\documentclass[aps,pra,twocolumn]{revtex4}
\usepackage{graphicx}
\usepackage{dcolumn}
\usepackage{bm}
\usepackage{amsmath}
\usepackage{amssymb,amsthm}
\usepackage{epsfig,color}
\usepackage[sans]{dsfont}

\definecolor{nblue}{rgb}{0.2,0.2,0.7}
\definecolor{ngreen}{rgb}{0.2,0.6,0.2}
\definecolor{nred}{rgb}{0.7,0.2,0.2}
\definecolor{nblack}{rgb}{0,0,0}

\def\bea{\begin{eqnarray}}
\def\eea{\end{eqnarray}}

\begin{document}

\title{Entanglement amplification of fermionic systems in an accelerated frame}

\author{Younghun Kwon}
\email{yyhkwon@hanyang.ac.kr}
\author{Jinho Chang}

\affiliation{Department of Physics, Hanyang University, Ansan,
Kyunggi-Do, 425-791, South Korea}
\date{\today}

\begin{abstract}
In this article we present an analysis to derive physical results in
the entanglement amplification of fermonic systems in the
relativistic regime, that is, beyond the single-mode approximation.
This leads a recent work in [M. Montero and E.
Mart\'{i}n-Mart\'{i}nez, JEHP 07 (2011) 006] to a physical result,
and solidifies that phenomenon of entanglement amplification can
actually happen in the relativistic regime.

\end{abstract}

\maketitle

\section{Introduction}


Entanglement, that is, correlations existing in multipartite quantum
systems, is an essential feature to describe information processing
of physical systems in the most fundamental level of physics. It is
known that, none of correlation measures which have been employed to
describe correlations of classical systems suffices to describe
entanglement. Then, a number of measures to quantify entanglement of
quantum systems in the non-relativistic regime have been developed,
for instance, the criteria based on the partial transpose has been
one of major tools to not only detect but also quantify entangled
states. The measure has also been used to describe evolution of
entanglement in time, under different settings of circumstances of
given systems.

It is then of fundamental interest, also natural, to extend the
analysis to the relativistic regime, that is, entanglement of
relativistic quantum systems. In particular, the case that one of
two parties sharing entangled states moves in a uniform acceleration
has been considered \cite{ref:alsing1} \cite{ref:ball}
\cite{ref:fuentes}. Interestingly, it turns out that entanglement
behaves differently according to physical systems, whether given
systems are bosonic or fermionic. One can contrast that, in the
relativistic regime, entanglement of bosonic systems disappears in
the limit of infinite acceleration, while entanglement of fermonic
systems shows a convergent behavior
\cite{ref:fuentes}\cite{ref:alsing2}. Among others, this is one of
the most remarkable features of quantum systems, that distinguish
physical systems in terms of entanglement behavior. Moreover, in the
recent, a counterintuitive phenomenon in entanglement of
relativistic quantum systems has been shown in Ref.
\cite{ref:montero1} that entanglement can be actually amplified when
one party is moving in a uniform acceleration. After these
qualitative results on entanglement of relativistic quantum systems
have been derived, the constraint to have physical results should be
taken into account, i.e. that detectors in relativistic quantum
systems is consistent to the behavior of entanglement in the
relativistic regime. In fact, this has been made, e.g. in the case
of entanglement behavior of fermonic systems in the infinite
acceleration \cite{ref:chang}.

In this paper, we make an analysis to derive physical results in the
entanglement amplification of fermonic systems in the relativistic
regime. We apply methods shown in Refs. \cite{ref:montero2}
\cite{ref:montero3} \cite{ref:chang} and provide physical results
that entanglement of fermionic systems can be amplified in an
accelerated frame. Our result not only makes a proper analysis
itself, but also solidifies the result that phenomenon of
entanglement amplification can actually happen (within the employed
measure).

The present article is organized as follows. In Sec. II, we will
give a brief description for fermionic system in an accelerated
frame. In Sec. III, we investigate the entanglement amplification of
pure and mixed quantum states in fermionic system. In Sec. IV, we
conclude and discuss our result.

\section{Accelerated Frame}

The accelerated frame can be described by the Rindler coordinate $(\tau,\varsigma,y,z)$ instead of Minkowski coordinate $(t,x,y,z)$. In right wedge of it(called region $I$), it can be described by $\displaystyle
ct=\varsigma \sinh(\frac{a \tau}{c}),x=\varsigma \cosh(\frac{a\tau}{c}) $ and in left one of it(region $II$) the coordinate is given by $ ct=-\varsigma \sinh(\frac{a \tau}{c}),x=-\varsigma \cosh(\frac{a \tau}{c})$, where $a$ denotes the fixed acceleration of the frame and $c$ is the velocity of light. For a fixed
$\varsigma$, the coordinate displays hyperbolic trajectories in space-time.

A field in Minkowski and Rindler space-time can be expressed as $
\displaystyle \phi = N_{M}\sum_{i}(a_{i,M}v^{+}_{i,M} + b^{\dag}_{i,M}v^{-}_{i,M} ) = N_{R}\sum_{j}(a_{j,I}v^{+}_{j,I} + b^{\dag}_{j,I}v^{-}_{j,I} + a_{j,II}v^{+}_{j,II} + b^{\dag}_{j,II}v^{-}_{j,II} ) $. Here $N_{M}$ and $N_{R}$ are the normalization constants. $v^{\pm}_{i,M}$ refers to the positive and negative energy solutions of the Dirac equation in Minkowski space-time, which can be obtained with
 respect to the Killing vector field in Minkowski space-time. $v^{\pm}_{i,I}$
 and $v^{\pm}_{i,II}$ are the positive and negative energy solutions of the
 Dirac equation in Rindler spacetime, with
 respect to the Killing vector field in region $I$ and $II$.
 Also  $a^{\dag}_{i,\lambda}(a_{i,\lambda})$ and
$b^{\dag}_{i,\lambda}(b_{i,\lambda})$ are the creation(annihilation)
operators, satisfying the anti-commutation relations,  for the
positive and negative energy solutions(particle and antiparticle),
where $\lambda $ denotes $M,I,II$. A combination of Minkowski mode,
called Unruh mode, can be transformed into monochromatic Rindler
mode and can annihilate the same Minkowski vacuum: $\displaystyle
A_{i,R/L}\equiv \cos \gamma_{i}a_{i,I/II} - \sin \gamma_{i}
b^{\dag}_{i,II/I}, $ where $\cos \gamma_{i}=(e^{\frac{-2 \pi \Omega
c}{a}}+1)^{-1/2}$.  More generally, we have $\displaystyle
a^{\dag}_{i,U}=q_{L}(A^{\dag}_{\Omega ,L}\otimes I_{R}) +
q_{R}(I_{L} \otimes  A^{\dag}_{\Omega ,R})$ beyond the single mode
approximation. Using this relation, in case of a Grassmann scalar,
the Unruh vacuum and the one-particle state are given by
\begin{eqnarray}
|0_{\Omega }\rangle_{U} &=& \cos^{2} \gamma_{\Omega }
|0000\rangle_{\Omega } - \sin \gamma_{\Omega }\cos \gamma_{\Omega }
|0011\rangle_{\Omega }
\nonumber\\
        &+& \sin \gamma_{\Omega }\cos \gamma_{\Omega } |1100\rangle_{\Omega } - \sin^{2}
\gamma_{\Omega } |1111\rangle_{\Omega } \nonumber\\
|1_{\Omega }\rangle^{+}_{U} &=& q_{R}(\cos \gamma_{\Omega }
|1000\rangle_{\Omega } - \sin \gamma_{\Omega } |1011\rangle_{\Omega
})
\nonumber\\
                        &+& q_{L}(\sin \gamma_{\Omega } |1101\rangle_{\Omega } +\cos
 \gamma_{\Omega } |0001\rangle_{\Omega }
\end{eqnarray}
Here we consider $q_{R}$ and $q_{L}$ as a real number and use the
notation $|pqmn\rangle_{\Omega } \equiv |p_{\Omega
}\rangle^{+}_{I}|q_{\Omega }\rangle^{-}_{II}  |m_{\Omega
}\rangle^{-}_{I} |n_{\Omega }\rangle^{+}_{II} $. Actually there is
another possibility of one particle state such as

\begin{eqnarray}
|1_{\Omega }\rangle^{-}_{U} &=& q_{L}(\cos \gamma_{\Omega }
|0100\rangle_{\Omega } - \sin \gamma_{\Omega } |0111\rangle_{\Omega
})
\nonumber\\
                        &+& q_{R}(\sin \gamma_{\Omega } |1110\rangle_{\Omega } +\cos
 \gamma_{\Omega } |0010\rangle_{\Omega }
\end{eqnarray}
From now on, for simplicity, the index $\Omega$ will be omitted. The
single-mode approximation corresponds to the case of $q_{R}=1$.
Recently it was shown that the physical ordering of the fermionic
system should be rearranged by the sequence of particle and
antiparticle in the separated
region\cite{ref:montero2}\cite{ref:montero3}\cite{ref:chang}. So in
this report, we
will use the physical structure for fermionic system.\\

\section{Entanglement amplification of quantum states in fermionic system}

\subsection{2 party pure entangled states }

At first let us consider pure entanglement between Alice and Bob.
Two parties share a pure entangled state in fermionic system and Bob
travels with a uniform acceleration. Then, the state shared is
described as
\begin{equation}
|\Phi^{+} (\alpha) \rangle = \cos \alpha |0\rangle_{M}|0\rangle_{U}
+ \sin \alpha |1\rangle_{M}|1^{+}\rangle_{U}. \label{eq6}
\end{equation}

Suppose that Bob's detector cannot distinguish between the particle
and the antiparticle. As it is explained, two regions $I$ and $II$
are causally disconnected and thus Bob does not have assess to both.
Hence, the state between Alice and Bob can be found by tracing
either regions. The state of Alice and Bob when Bob is in region $I$
is described as,
\begin{eqnarray}
\rho_{AB_{\mathrm{I}}}^{\Phi^{+}}
&=& \cos ^2 \alpha \cos ^4 \gamma |000\rangle \langle 000| \nonumber\\
&+& \frac{q_{R}}{2} \sin 2\alpha \cos ^{3} \gamma (|000\rangle \langle 110| + |110\rangle \langle 000|) \nonumber\\
&+& q_{L}^{2} \sin ^{2} \alpha \cos ^{2} \gamma |100\rangle \langle 100| \nonumber\\
&+& \frac{1}{2}(1-(1-2 q_{L}^{2}) \cos 2\gamma) \sin ^{2}\alpha
|110\rangle \langle 110| \nonumber\\
&-&\frac{q_{L}}{2}  \sin 2\alpha \cos ^{2}\gamma \sin
\gamma(|001\rangle \langle 100| + |100\rangle
\langle 001|) \nonumber\\
&-& \frac{q_{R}q_{L}}{2} \sin ^{2}\alpha \sin 2\gamma (|100\rangle
\langle 111| + |111\rangle \langle 100| ) \nonumber\\
&+& \frac{1}{4} \cos ^{2}\alpha  \sin ^{2} 2\gamma (|001\rangle
\langle 001| + |010\rangle \langle 010|) \nonumber\\
&+& \frac{q_{R}}{2} \sin 2\alpha \cos \gamma \sin ^{2} \gamma
(|001\rangle \langle 111| + |111\rangle \langle 001|) \nonumber\\
&+& q_{R}^{2} \sin ^{2} \alpha \sin ^{2} \gamma |111\rangle \langle 111| \nonumber\\
&+& \frac{q_{L}}{2} \sin 2\alpha \sin ^{3} \gamma (|011\rangle \langle 110| + |110\rangle \langle 011|) \nonumber\\
&+& \cos ^{2} \alpha \sin ^{4} \gamma |011\rangle \langle 011|.
\nonumber
\end{eqnarray}

\begin{figure}[h!]
\begin{center}
\includegraphics[width=6cm]{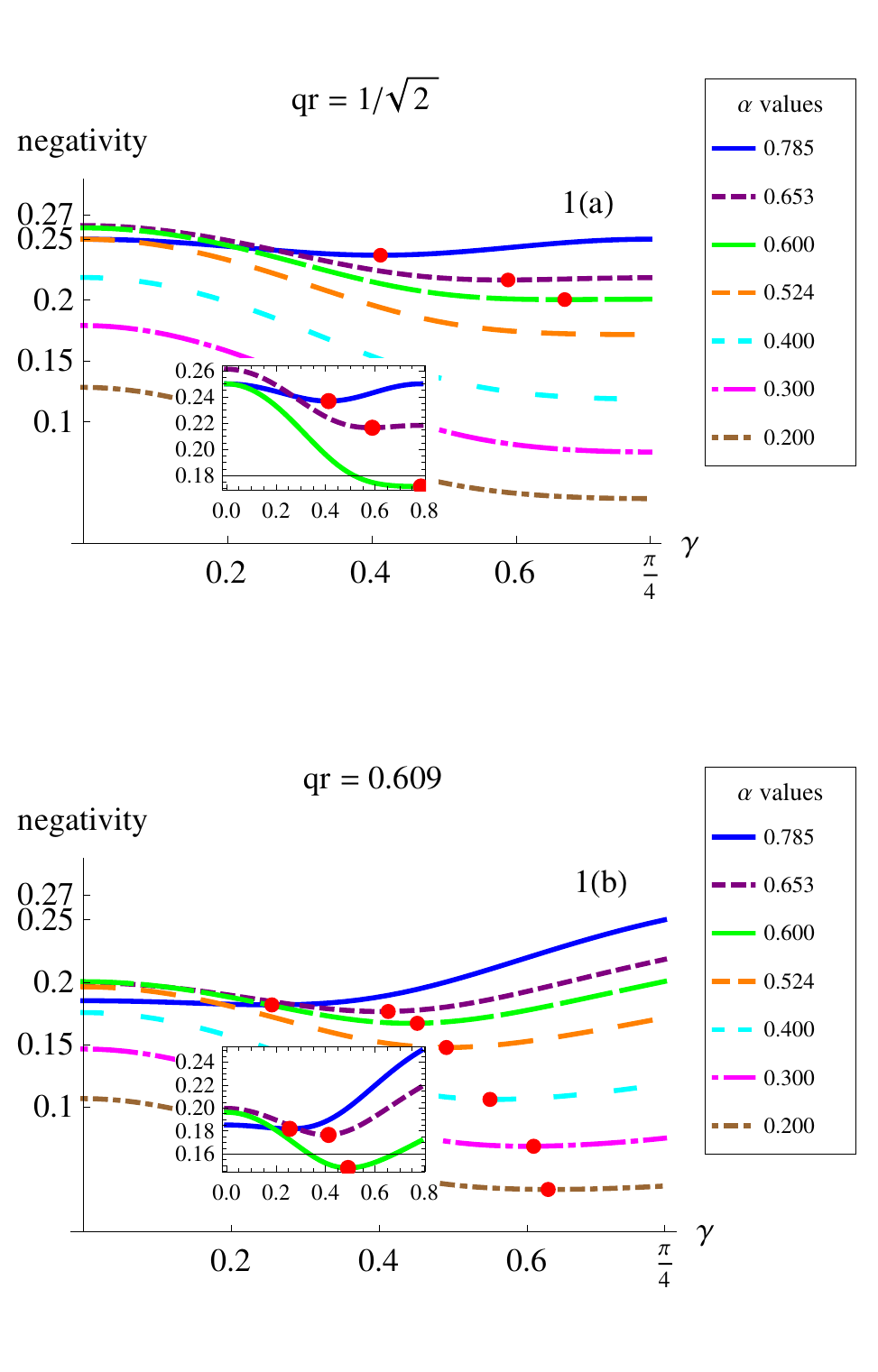}
\caption{\label{fig2}(Color online) Negativity of quantum state $
\rho_{AB_{\mathrm{I}}}^{\Phi^{+}}$. Part (a) and (b) show the cases
of $q_{R}=\frac{1}{\sqrt{2}}$ and $q_{R}=0.609$ respectively. The
dot-point on the line denotes the point of variation of the curve.
The small box inside the figure displays the negativity of quantum
states with $\alpha=0.785$,$\alpha=0.653$ and $\alpha=0.600$ for $
\rho_{AB_{\mathrm{I}}}^{\Phi_{+}}$. Some quantum states reveals
clearly the entanglement amplification as acceleration increases.
Here $\gamma=\frac{\pi}{4}$ denotes the infinite acceleration. }
\end{center}
\end{figure}

 The entanglement of quantum state is measured by negativity, which
evaluates the sum of negative eigenvalues of the partial transposed
density matrix. The entanglement behavior at
$q_{R}=\frac{1}{\sqrt{2}}$ is depicted in Fig. 1(a). In fact similar
analysis was reported in \cite{ref:montero2} and
\cite{ref:montero4}. The entanglement of the quantum state is
decreased to the dot-point on the line. However after that point the
entanglement of the state begins to increase, as Bob moves in more
accelerated frame. It means the entanglement amplification of the
quantum state in terms of acceleration. It is very surprising
phenomena, since it is commonly believed that the entanglement may
not be generated by acceleration. Fig.1 shows that the entanglement
of some quantum states in fermionic system violates the common
belief. As seen in Fig. 1, lesser entangled the initial state is,to
more right position the point of variation moves. For the state
$\Phi^{+}(\alpha)$ the increase of entanglement happens up to
$\alpha=0.523599$ when $q_{R}=\frac{1}{\sqrt{2}}$. That is, the
amplification of entanglement can be clearly seen between
$\frac{\pi}{4}\ge \alpha \ge 0.523599$, when
$q_{R}=\frac{1}{\sqrt{2}}$. The entanglement amplification at
$q_{R}=0.609$ can be found in Fig. 1(b). In this case some quantum states in the accelerated frame reveals more entanglement than that in an inertial one.\\

We next consider the entanglement between Alice and Bob, when they
share the following state,
\begin{equation}
|\Phi^{-} (\alpha) \rangle = \cos \alpha |0\rangle_{M}|0\rangle_{U}
+ \sin \alpha |1\rangle_{M}|1^{-}\rangle_{U} \label{eq7}.
\end{equation}
As it is explained previously, Bob has inaccessible part due to his
acceleration. The state of Alice and Bob after tracing the region
$II$ can be found. Actually as far as entanglement is concerned, the
entanglement behavior of $\Phi^{+}(\alpha)$ seems to be equivalent
to that of $\Phi^{-}(\alpha)$, which was discussed in \cite{ref:montero5}.\\
We now consider pure entangled states such as Eq. (5), when Bob is
traveling with a uniform acceleration, as follows:
\begin{equation}
|\Phi^{*} (\alpha) \rangle = \cos \alpha
|0\rangle_{M}|1^{+}\rangle_{U} + \sin \alpha
|1\rangle_{M}|0\rangle_{U}. \label{eq8}
\end{equation}
As it is done before, the state that Alice and Bob share can be
obtained beyond the single-mode approximation. The state when Bob is
in region $I$ is obtained by tracing the other region,
\begin{eqnarray}
\rho_{AB_{\mathrm{I}}}^{\Phi^{*}}
&=&  q_{L}^{2} \cos ^2 \alpha \cos ^2 \gamma |000\rangle \langle 000| \nonumber\\
&+& \frac{1}{2}(1-(1-2 q_{L}^{2})\cos 2\gamma) \cos ^{2} \alpha |010\rangle \langle 010| \nonumber\\
&+& \frac{q_{R}}{2}\cos ^3\gamma \sin 2 \alpha  |010\rangle \langle 100|+|100\rangle \langle 010|)  \nonumber\\
&+& \cos ^4\gamma \sin ^2 \alpha  |100\rangle \langle 100|  \nonumber\\
&-& \frac{q_{R}q_{L}}{2} \sin ^{2}\alpha \sin 2\gamma (|000\rangle
\langle 011| + |011\rangle \langle 000| ) \nonumber\\
&-& \frac{q_{L}}{2} \sin 2\alpha \cos ^2\gamma \sin
\gamma(|000\rangle\langle 101| + |101\rangle \langle 000| ) \nonumber\\
&+& q_{R}^{2} \cos ^{2} \alpha \sin ^{2} \gamma |011\rangle \langle 011| \nonumber\\
&+& \frac{q_{R}}{2} \sin 2\alpha \cos \gamma \sin
^2\gamma(|011\rangle \langle 101| + |101\rangle \langle 011| ) \nonumber\\
&+& \frac{1}{4} \sin ^2\alpha \sin
^2 2\gamma(|101\rangle \langle 101| + |110\rangle \langle 110| ) \nonumber\\
&+& \frac{q_{L}}{2}\sin ^3\gamma \sin 2 \alpha ( |010\rangle \langle 111|+|111\rangle \langle 010|)  \nonumber\\
&+& \sin ^{2} \alpha \sin ^{4} \gamma |111\rangle \langle 111|.
\nonumber
\end{eqnarray}

\begin{figure}[h!]
\begin{center}
\includegraphics[width=6cm]{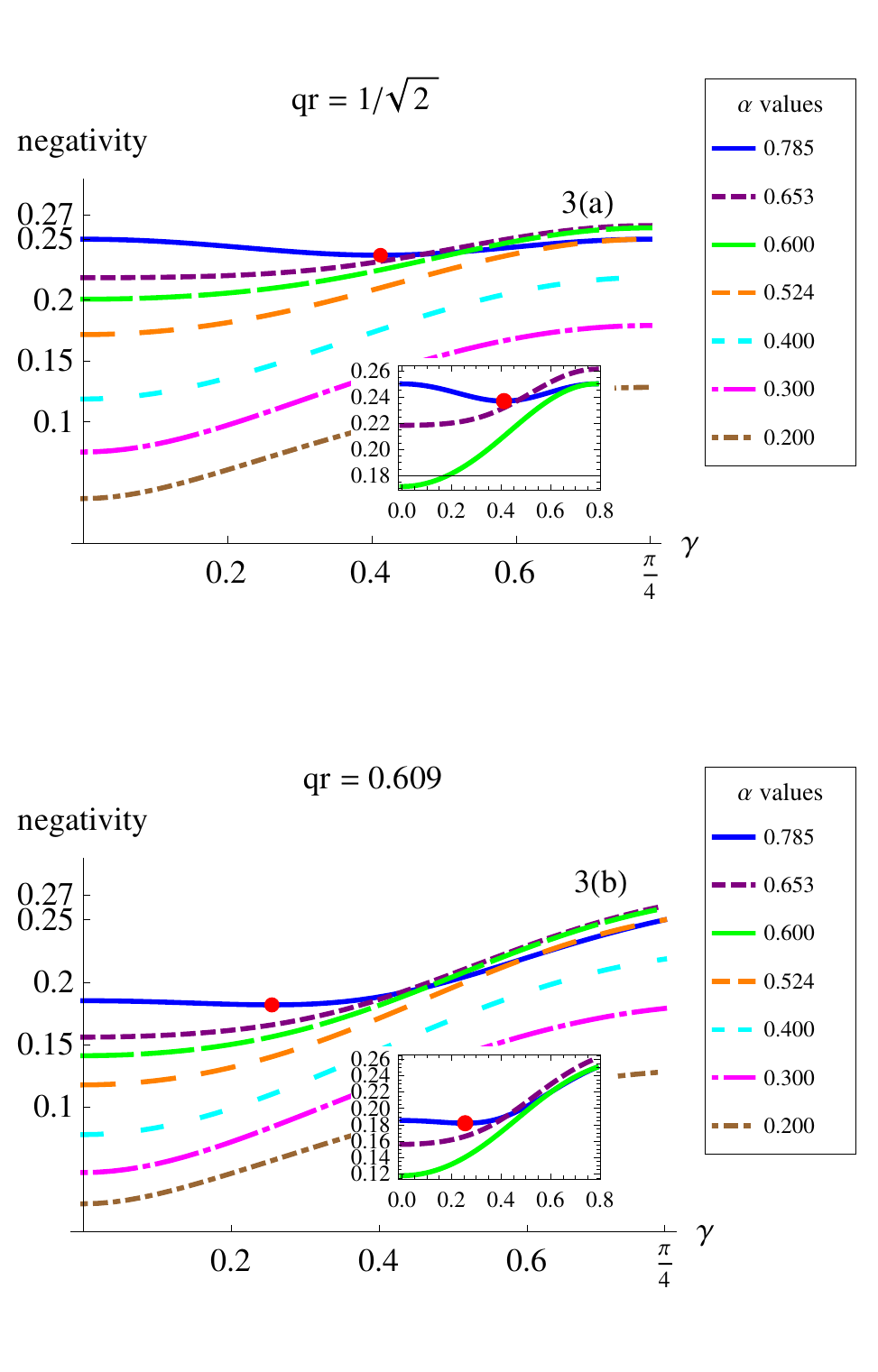}
\caption{\label{fig2}(Color online) Negativity of quantum state $
\rho_{AB_{\mathrm{I}}}^{\Phi^{*}}$. Part (a) and (b) show the cases
of $q_{R}=\frac{1}{\sqrt{2}}$ and $q_{R}=0.609$ respectively. The
dot-point on the line denotes the point of variation of the curve.
The small box inside the figure displays the negativity of quantum
states with $\alpha=0.785$,$\alpha=0.653$ and $\alpha=0.600$ for $
\rho_{AB_{\mathrm{I}}}^{\Phi^{*}}$. Some quantum states reveal
clearly the entanglement amplification as acceleration increases.
Here $\gamma=\frac{\pi}{4}$ denotes the infinite acceleration. }
\end{center}
\end{figure}

The entanglement behavior for the quantum state $\Phi^{*}(\alpha)$
at $q_{R}=\frac{1}{\sqrt{2}}$ can be seen in Fig. 2(a). We observe
the point of variation as well as the amplification of entanglement,
for certain quantum states. However, its entanglement behaves
differently from that of state $\Phi^{+}(\alpha)$. The main
difference is that the entanglement for some range of $\alpha$ does
not decrease rather increases as Bob's acceleration is getting
larger. For example, the quantum state with $\alpha=0.653$ at
$q_{R}=\frac{1}{\sqrt{2}}$, can get the maximal entanglement at the
infinite acceleration. That is, the entanglement of the quantum
state with specific $q_{R}$ and $\alpha$ has the largest value at
the infinite acceleration. It seems to be a strange property, since
the acceleration is believed to reduce the entanglement of the
quantum state. At $q_{R}=0.609$, the behavior of
entanglement can be found in Fig. 2(b).\\

\subsection*{III.b 2 party mixed entangled state }

Up to now, we have considered entanglement of pure states in
fermionic system when one of parties is traveling with a uniform
acceleration. We have observed that there is an amplification of
entanglement when a partner sharing a pure quantum state moves in
accelerated frame. In this subsection, we consider a more complicate
scenario when two parties share a mixed state. It is aimed to find
how the entanglement behavior depends on the mixedness property, and
also if its amplification is related to the mixedness. In
particular, the case when a white noise is added to a maximally
entangled states, so-called Werner state, is to be considered. The
mixedness of Werner states is parameterized by a single parameter.
So we suppose that two parties Alice and Bob prepare Werner states
in inertial frames, and then Bob moves in the uniformly accelerated
frame. That is, the initial state of Alice and Bob can be expressed
as follows,
\begin{equation}
\rho_{W}= F |\Phi_{+} (\alpha=\pi/4) \rangle  \langle \Phi_{+}
(\alpha=\pi/4) | + \frac{1-F}{4}\mathbb{I}\label{eq9},
\end{equation}
where the maximally entangled state is taken from Eq. (\ref{eq6})
when $\alpha=\pi/4$.

Suppose that Bob moves in an accelerated frame. Beyond the
single-mode approximation, the state that Alice and Bob share in
Bob's region $I$ is obtained by tracing the region $II$, as follows,

\begin{widetext}
\begin{eqnarray}
\rho_{AB_{\mathrm{I}}}^{W} &=& \frac{1}{2}F q_{R} \cos ^3 \gamma
(|000\rangle \langle 110|+|110\rangle \langle 000|) + \frac{1}{8} \cos ^{2} \gamma (3-2 q_{R}^{2}+F(1-2q_{R}^{2})+(1-F)\cos 2\gamma)|100\rangle \langle 100| \nonumber\\
&+& \frac{1}{8} \cos ^{2} \gamma (3-2q_{R}^{2}-F(1-2q_{R}^{2})+(1+F)\cos 2\gamma)|000\rangle \langle 000| -\frac{F q_{L}}{2} \cos ^{2}\gamma \sin \gamma(|001\rangle \langle 100| + |100\rangle \langle 001|) \nonumber\\
&+&\frac{F q_{R}}{2} \cos \gamma \sin ^{2}\gamma(|001\rangle \langle 111| + |111\rangle \langle 001|) +\frac{F q_{L}}{2} \sin ^{3}\gamma(|011\rangle \langle 110| + |110\rangle \langle 011|) \nonumber\\
&+& \frac{1}{4} \sin ^{2} \gamma ((1+F)q_{R}^{2}+(1-F) \sin ^{2}\gamma)|111\rangle \langle 111| + \frac{1}{4} \sin ^{2} \gamma ((1-F)q_{R}^{2}+(1+F) \sin ^{2}\gamma)|011\rangle \langle 011| \nonumber\\
&-& \frac{1}{8} (1-F)q_{L}q_{R} \sin 2 \gamma (|000\rangle \langle 011| +|011\rangle \langle 000|) - \frac{1}{8} (1+F)q_{L}q_{R} \sin 2 \gamma (|100\rangle \langle 111| +|111\rangle \langle 100|)\nonumber\\
&+&\frac{1}{16} (1-F) \sin ^{2} 2\gamma|101\rangle \langle 101|+\frac{1}{16} (1+F) \sin ^{2} 2\gamma|001\rangle \langle 001| +\frac{1}{16} (2(1+F)-2(1+F)(1-2 q_{R}^{2}) \cos 2\gamma \nonumber\\
& &+(1-F)\sin ^{2} 2\gamma)|110\rangle \langle 110| +\frac{1}{16}
(2(1-F)-2(1-F)(1-2 q_{R}^{2}) \cos 2\gamma +(1+F)\sin ^{2} 2\gamma)
|010\rangle \langle 010| \nonumber
\end{eqnarray}
\end{widetext}

\begin{figure}[h!]
\begin{center}
\includegraphics[width=6cm]{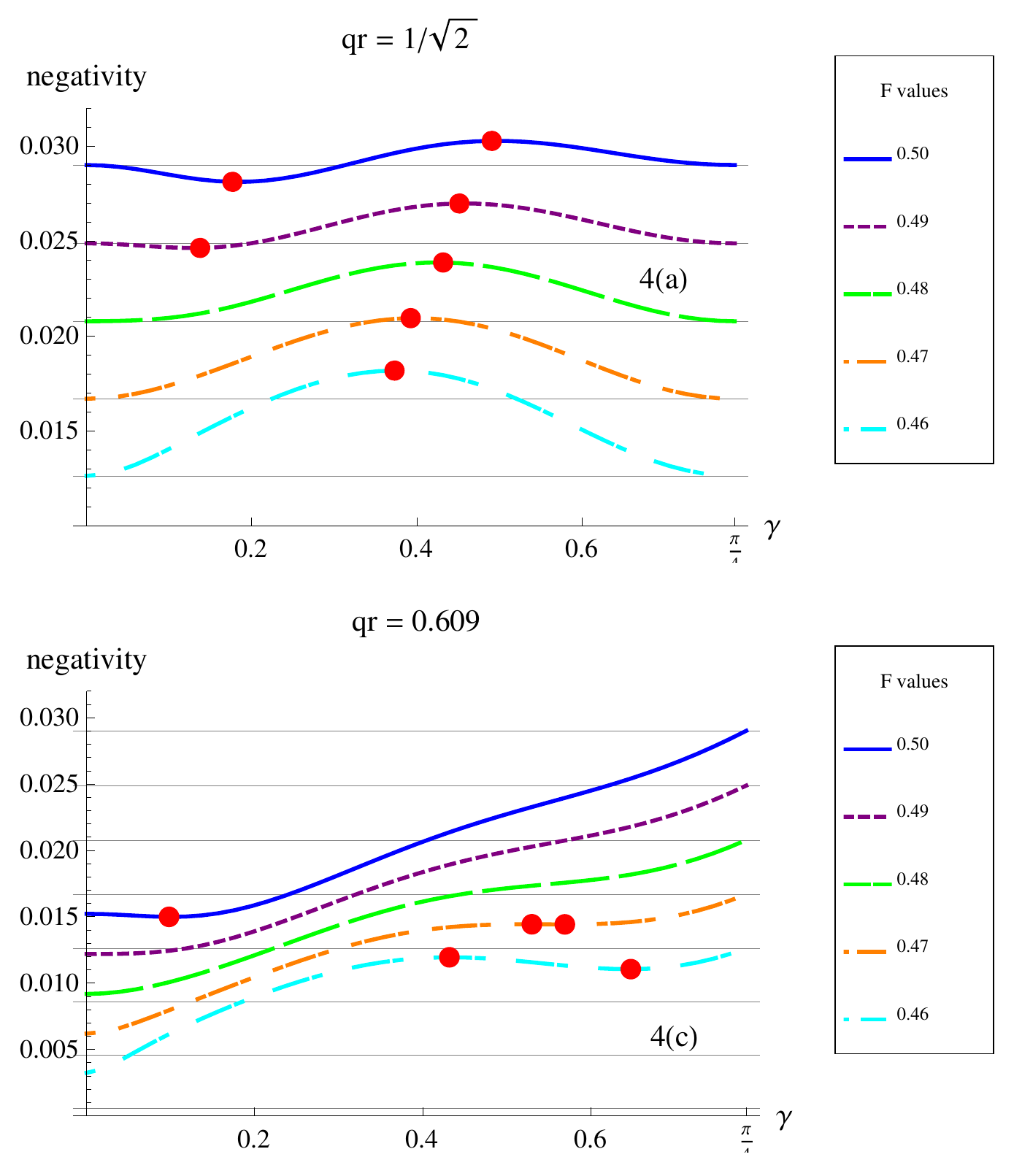}
\caption{\label{fig3}(Color online) Negativity of quantum state $
\rho_{AB_{\mathrm{I}}}^{W}$. Part (a) and (b) show the cases of
$q_{R}=\frac{1}{\sqrt{2}}$ and $q_{R}=0.609$ respectively. The
dot-point on the line denotes the point of variation of the curve.
Some quantum states display two dot-points, implying that there are
two regions for entanglement behavior. Part (a) and (b) show clearly
the entanglement amplification as acceleration increases. Here
$\gamma=\frac{\pi}{4}$ denotes the infinite acceleration. }
\end{center}
\end{figure}

The negativity in Fig.3 depicts the entanglement amplification for
some Werner states. It implies that when the mixed states are
considered, their entanglement for certain mixed states shows the
amplification behavior. Specially the point of variation can be
found two times for the Werner state with $F=0.50$ at
$q_{R}=\frac{1}{\sqrt{2}}$. In fact it happens in the Wener state of
$F=0.50$ at $q_{R}=\frac{1}{\sqrt{2}}$ or $F=0.49$ at
$q_{R}=\frac{1}{\sqrt{2}}$ or $F=0.47$ at $q_{R}=0.609$ or $F=0.46$
at $q_{R}=0.609$. It means that there are two regions of
entanglement amplification for the states, which seems to be a
peculiar property of mixed states.\\
We now consider the mixed entangled states such as Eq. (7), when Bob
is traveling with a uniform acceleration, as follows:
\begin{eqnarray}
\rho_{WL} &=& F |\Phi_{+} (\alpha=\pi/4) \rangle  \langle \Phi_{+}
(\alpha=\pi/4) | \nonumber\\
&+& \frac{1-F}{2}(|01 \rangle  \langle 01 |+|10 \rangle  \langle 10
|)\label{eq10},
\end{eqnarray}
where the maximally entangled state is taken from Eq. (\ref{eq6})
when $\alpha=\pi/4$. As it is done before, the state that Alice and
Bob share can be obtained beyond the single-mode approximation. The
state when Bob is in region $I$ is obtained by tracing the other
region,

\begin{eqnarray}
& &\rho_{AB_{\mathrm{I}}}^{WL} = \frac{1}{2} F q_{R} \cos ^3 \gamma
(|000\rangle \langle 110|+|110\rangle \langle 000|)\nonumber\\
&+&\frac{1}{2}
\cos\gamma^2(F(q_{L}^2)+(1-F)\cos\gamma^2)|100\rangle\langle100|\nonumber\\
&+&\frac{1}{2}\cos\gamma^2((1-F)(q_{L}^2)+F\cos\gamma^2)|000\rangle\langle000|\nonumber\\
&+&\frac{1}{16}(1+3F-4F(1-2q_{R}^2)\cos2\gamma-(1-F)\cos4\gamma)|110\rangle\langle110|\nonumber\\
&-&\frac{1}{2}F q_{L}
\cos\gamma^2\sin\gamma(|001\rangle\langle100|+|100\rangle\langle001|)\nonumber\\
&+&\frac{1}{2}F q_{R} \cos\gamma
\sin\gamma^2(|001\rangle\langle111|+|111\rangle\langle001|)\nonumber\\
&+&\frac{1}{2}F q_{L}
\sin\gamma^3(|011\rangle\langle110|+|110\rangle\langle011|)\nonumber\\
&+&\frac{1}{2}\sin\gamma^2((1-F)q_{R}^2+F\sin\gamma^2)|011\rangle\langle011|\nonumber\\
&+&\frac{1}{2}(F q_{R}^2
\sin\gamma^2+(1-F)\sin\gamma^4)|111\rangle\langle111|\nonumber\\
&-&\frac{1}{4}(1-F)q_{R}q_{L}\sin2\gamma
(|000\rangle\langle011|+|011\rangle\langle000|)\nonumber\\
&-&\frac{1}{4}F
q_{R}q_{L}\sin2\gamma(|100\rangle\langle111|+|111\rangle\langle100|)\nonumber\\
&+&\frac{1}{8}(1-F)\sin2\gamma^2|101\rangle\langle101|\nonumber\\
&+&\frac{1}{8}F\sin2\gamma^2|001\rangle\langle001|\nonumber\\
&+&\frac{1}{8}(2(1-F)(1-(1-2q_{R}^2)\cos2\gamma)+F\sin2\gamma^2)|010\rangle\langle010|\nonumber
\end{eqnarray}

\begin{figure}[h!]
\begin{center}
\includegraphics[width=6cm]{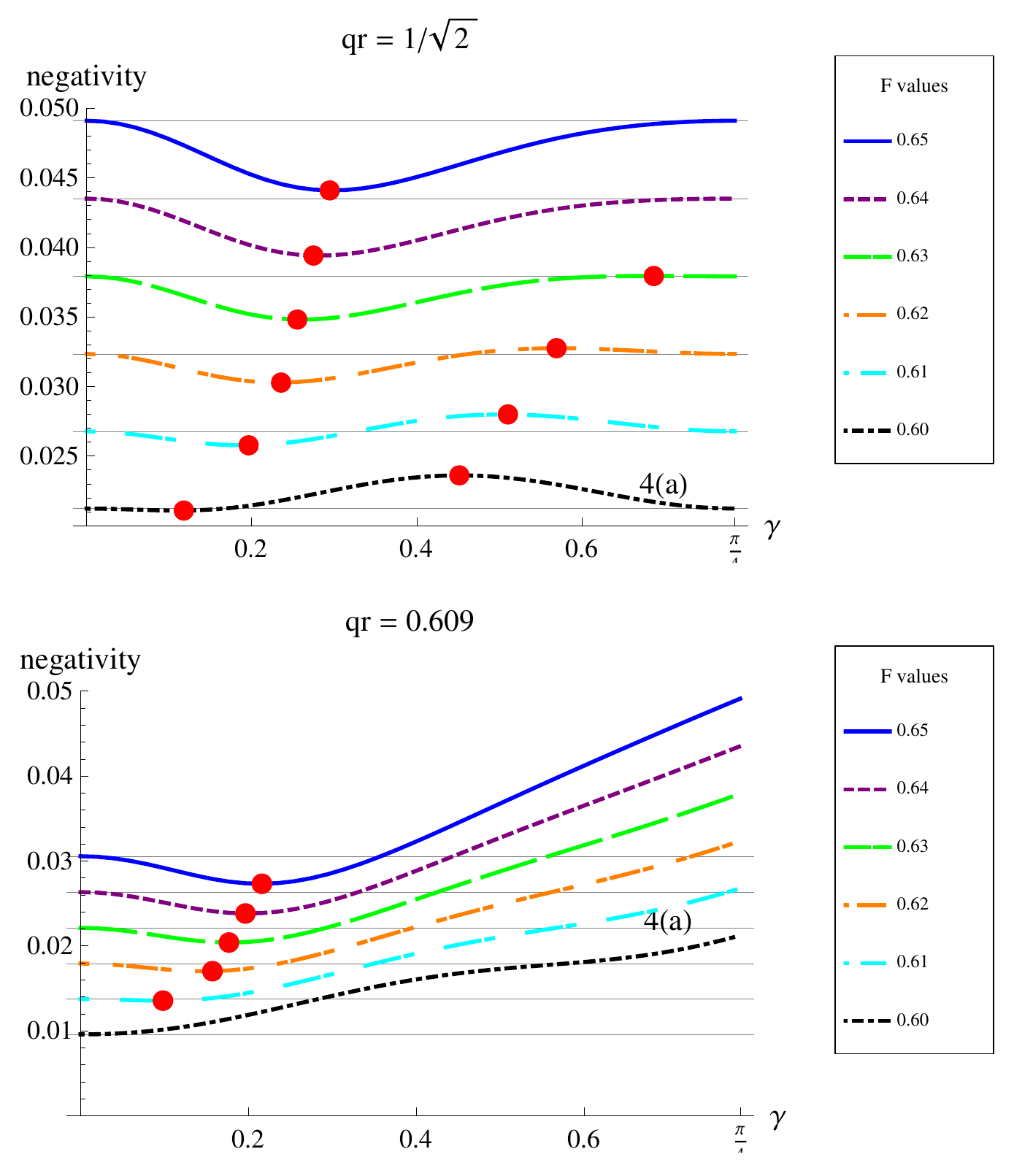}
\caption{\label{fig4}(Color online) Negativity of quantum state $
\rho_{AB_{\mathrm{I}}}^{WL}$. Part (a) and (b) show the cases of
$q_{R}=\frac{1}{\sqrt{2}}$ and $q_{R}=0.609$ respectively. The
dot-point on the line denotes the point of variation of the curve.
Some quantum states show clearly the entanglement amplification as
acceleration increases. Here $\gamma=\frac{\pi}{4}$ denotes the
infinite acceleration. }
\end{center}
\end{figure}

The negativity in Fig.4 shows the entanglement amplification for
some mixed quantum states of Eq. (\ref{eq10}). It implies that when
the mixed states are considered, their entanglement for certain
mixed states shows the amplification behavior. Specially the point
of variation can be found two times for $\rho_{WL}$  with $F=0.63$
at $q_{R}=\frac{1}{\sqrt{2}}$. In fact it happens in $\rho_{WL}$ of
$F=0.63$ at $q_{R}=\frac{1}{\sqrt{2}}$ or $F=0.62$ at
$q_{R}=\frac{1}{\sqrt{2}}$ or $F=0.61$ at $q_{R}=\frac{1}{\sqrt{2}}$
or $F=0.60$ at $q_{R}=\frac{1}{\sqrt{2}}$.

\section*{IV. Discussion and Conclusion } We have investigated
the amplification of entanglement of quantum states in fermionic
system when a party sharing entangled quantum state travels in
uniformly accelerated frame. Even though it has been widely believed
that the acceleration may spoil the entanglement of the system, we
showed that there can be the amplification of entanglement for some
quantum states regardless of pure or mixed one. Also it is a
surprise that some mixed states reveal two points of variation in
the line of negativity. It seems to be worthwhile to investigate why
there are more than one region of amplification of entanglement for
some mixed states.

\section*{Acknowledgment}
We would like to thank Dr. Mart\'{i}n-Mart\'{i}nez for pointing ref
\cite{ref:montero4} and Dr. Joonwoo Bae for a careful reading of the
manuscript and for valuable comments. This work is supported by
Basic Science Research Program through the National Research
Foundation of Korea funded by the Ministry of Education, Science and
Technology (KRF2011-0027142).

\end{document}